\documentclass[pre,showpacs]{revtex4}
\usepackage[colorlinks,citecolor=blue,linkcolor=red,dvipdfm]{hyperref}
\usepackage{amssymb}
\usepackage[dvips]{graphicx}
\usepackage{amsmath}
\usepackage{graphicx}
\usepackage{makeidx}
\usepackage{subfigure}

\begin{document}

\title{Bright solitons in defocusing media with spatial modulation of the
quintic nonlinearity}
\author{Jianhua Zeng$^{1}$ and Boris A. Malomed$^{2}$}
\affiliation{$^{1}$State Key Laboratory of Low Dimensional Quantum Physics, Department of
Physics, Tsinghua University, Beijing 100084, China\\
$^{2}$Department of Physical Electronics, School of Electrical Engineering,
Faculty of Engineering, Tel Aviv University, Tel Aviv 69978, Israel}

\begin{abstract}
It has been recently demonstrated that self-defocusing (SDF) media with the
cubic nonlinearity, whose local coefficient grows from the center to
periphery fast enough, support stable bright solitons, without the use of
any linear potential. Our objective is to test the genericity of this
mechanism for other nonlinearities, by applying it to one- and
two-dimensional (1D and 2D) quintic SDF media. The models may be implemented
in optics (in particular, in colloidal suspensions of nanoparticles), and
the 1D model may be applied to the description of the Tonks-Girardeau gas of
ultracold bosons. In 1D, the nonlinearity-modulation function is taken as $%
g_{0}+\sinh ^{2}\left( \beta x\right) $. This model admits a subfamily of
exact solutions for fundamental solitons. Generic \ soliton solutions are
constructed in a numerical form, and also by means of the Thomas-Fermi and
variational approximations (TFA and VA). In particular, a new ansatz for the
VA is proposed, in the form of ``raised $\mathrm{sech}$", which provides for
an essentially better accuracy than the usual Gaussian ansatz. The stability
of all the fundamental (nodeless) 1D solitons is established through the
computation of the corresponding eigenvalues for small perturbations, and
also verified by direct simulations. Higher-order 1D solitons with two nodes
have a limited stability region, all the modes with more than two nodes
being unstable. It is concluded that the recently proposed ``anti-VK"
stability criterion for fundamental bright solitons in systems with SDF
nonlinearities holds here too. Particular exact solutions for 2D solitons
are produced as well.
\end{abstract}

\maketitle

\section{Introduction}

The guiding of matter waves in Bose-Einstein condensates (BECs) \cite{ref1}
and light in optical waveguides \cite{PhotCryst}, using effective potentials
which are induced by optical lattices or photonic crystals, has drawn a
great deal of attention in recent years. Such settings play a crucial role
in the creation and stabilization of ordinary and gap solitons \cite%
{ref1,ref3, ref4}, supported by the balance of the diffraction,
lattice-induced potentials, and self-focusing or defocusing nonlinearity,
respectively.

In addition to linear lattices, many works have been dealing with their
nonlinear counterparts (the spatial modulation of the local nonlinearity
coefficient) \cite{ref4}-\cite{ref14} and combined linear-nonlinear \cite%
{antiVK}-\cite{ref12} lattices. The nonlinear lattices may be realized,
severally in optics and BECs, by means of properly designed
photonic-crystals structures, or nonlinearity landscapes induced by the
Feshbach resonance controlled by nonuniform external fields. A recent
noteworthy result is the existence of stable bright solitons in
self-defocusing (SDF) $D$-dimensional media with the strength of the cubic
nonlinearity growing toward the periphery, as a function of radius $r$, at
any rate faster than $r^{D}$ \cite{ref13,ref13-2}.

Along with the cubic nonlinearity, quintic terms appear in various physical
settings too. In optics, the quintic nonlinearity was theoretically
predicted \cite{ref16, ref17} and experimentally observed in fluids \cite%
{ref18, ref19} and glasses \cite{ref20}-\cite{ref24}. In the context of BEC,
the quintic nonlinearity represents three-body interactions in a dense
condensate, provided that collision-induced losses may be neglected \cite%
{ref15}.
The self-focusing quintic nonlinearity is critical in 1D, similar to its
cubic counterpart in 2D case \cite{ref5}, i.e., it gives rise to a
degenerate family of \textit{Townes solitons }\cite{ref6}, with a single
value of the norm for the entire family. In the free space, the Townes
solitons are unstable, but they can be readily stabilized by linear
potentials, as demonstrated in 1D \cite{ref25}-\cite{ref27} and 2D \cite%
{BBB,ref3} settings alike. The use of nonlinear potentials for the same
purpose turns out to be tricky: in the 2D space with the cubic
self-focusing, only nonlinearity-modulation profiles with \emph{sharp edges}
are able to stabilize the Townes solitons in a limited parameter region \cite%
{sharp,ref4}. In the 1D system with the quintic nonlinearity, nonlinear
lattices with a \emph{smooth} (sinusoidal) modulation profile stabilize the
respective Townes solitons against the critical collapse, but only in a
narrow region \cite{MIW-quintic,ref14}.

On the other hand, in Ref. \cite{Kolo} it was demonstrated that the 1D
Tonks-Girardeau (TG) gas of bosons with the hard-core repulsion, emulating a
degenerate Fermi gas, obeys the 1D nonlinear Schr\"{o}dinger (NLS) equation
with the quintic SDF term (an additional nonlocal cubic interaction appears
if the bosons carry dipole moments \cite{BBB-TG}). While this equation is
inappropriate for the description of dynamical effects caused by
interference of bosonic wavepackets \cite{Gir}, it correctly predicts
stationary patterns in the trapped gas \cite{Kolo,Gir,BBB-TG}.
Experimentally, the TG gas was realized in an ultracold gas of $^{87}$Rb
loaded into a tightly confined potential pipe \cite{experTG}. In fact, the
``quantum Newton's cradle", realized in a trapped chain of $%
^{87}$Rb atoms \cite{experTG2}, is another example of the TG gas.

In this work, we address the possibilities to support stable bright solitons
in 1D and 2D media with the SDF \textit{quintic} term whose local strength
grows fast enough at $r\rightarrow \infty $ (in fact, the solitons exist for
the growth rates faster than $r^{2D}$). These results generalize those
recently reported in Refs. \cite{ref13} and \cite{ref13-2} for SDF media
with the modulated cubic nonlinearity, and thus demonstrate the genericity
of the method for creating robust bright solitons using solely the SDF
nonlinearity. We focus on 1D models with the modulation profiles in the form
of hyperbolic functions, see Eqs. (\ref{g}) and (\ref{g2D}) below, for which
some soliton solutions can be obtained in an exact analytical form, and
generic ones are constructed numerically. The family of the fundamental
solitons is completely stable, similar to the situation reported in Refs.
\cite{ref13} and \cite{ref13-2} in the case of the spatially modulated cubic
SDF nonlinearity. We give an additional argument, based on energy estimates,
in favor of the conjecture that the fundamental solitons may realize the
system's ground state. In terms of the dependence between the soliton's
propagation constant and norm, $k(N)$, the bright solitons in the SDF medium
obey the \textit{inverted} Vakhitov-Kolokolov (alias \textit{anti-VK} \cite%
{antiVK}) criterion, $dk/d{N}<0$, on the contrary to the usual VK criterion
for fundamental solitons in self-focusing media, $dk/dN>0$ \cite{ref5,VK}.
It appears that, as conjectured in Ref. \cite{antiVK}, the anti-VK criterion
is a necessary but, in the general case, not sufficient condition for the
stability of bright solitons supported by SDF nonlinearities. Higher-order
1D solitons, whose shapes feature different numbers of nodes ($n$), are
found too. A part of the soliton family with $n=2$ is stable, while all the
modes with $n\geq 3$ are found to be unstable.

It is relevant to mention that the model considered here, as well as its
previously studied counterpart with the cubic nonlinearity \cite%
{ref13,ref13-2}, is not integrable. Indeed, it was proven (see,
e.g., Ref. \cite{nonintegrable}) that the NLS equation in 1D with
any nonlinearity different from pure cubic is nonintegrable, hence
our quintic model is definitely nonintegrable too. For this reason,
the localized modes, that we study in this paper, are not
``solitons" in the rigorous mathematical meaning of the word, but
rather ``solitary waves". Nevertheless, we call the modes
``solitons", following the usage commonly adopted in the current
literature. In this work, we do not study interactions between the
solitons, as multi-soliton configurations trapped in the nonlinear
potential well are less relevant than single-soliton states.

The rest of the paper is organized as follows. In Section II, we introduce
the model, report particular exact solutions, and develop the variational
approximation (VA) for generic 1D solitons, based on Gaussian and sech
\textit{ans\"{a}tze} (the latter one is taken in a generalized form, based
on ``raised sech", which is an essential technical novelty). Numerical
results for the 1D fundamental and higher-order solitons, including the
stability analysis, are reported in Section III. 2D solitons are briefly
considered in Section IV, and the paper is concluded by Section V.

\section{The quintic model and analytical approximations}

\subsection{The model and exact solutions}

The NLS equation with the quintic SDF nonlinearity for field amplitude $u(%
\mathbf{r},z)$ is
\begin{equation}
iu_{z}=-\frac{1}{2}\nabla ^{2}u+[g_{0}+g(\mathbf{r)}]|u|^{4}u,  \label{GPE}
\end{equation}%
where $g_{0}\geq 0$ is the strength of the uniform quintic term, Laplacian $%
\nabla ^{2}=\partial _{x}^{2}+\partial _{y}^{2}$ or $\nabla ^{2}=\partial
_{x}^{2}$ acts on transverse coordinates $\mathbf{r}=\left( x,y\right) $ or $%
x$ in 2D and 1D settings, respectively, $z$ is the propagation coordinate in
the optical medium (in the Gross-Pitaevskii equation for matter waves, $z$
is replaced by time $t$), and positive function $g(r)$ accounts for the
nonlinearity strength growing at $r\rightarrow \infty $. In optics, the 1D
version of this modulated nonlinearity may be implemented in a planar
waveguide of a variable transverse width, filled with a colloidal suspension
of metallic nanoparticles, as proposed in Ref. \cite{ref14}. The same 1D
model applies to the TG gas trapped in a potential pipe whose confinement
strength varies along the axial coordinate, similar to how it was recently
proposed to induce a modulated cubic nonlinearity in the effectively
one-dimensional BEC with attractive inter-atomic interactions \cite{Luca}.

Stationary solutions to Eq. (\ref{GPE}) with propagation constant $k$ are
sought for as $u(\mathbf{r},z)=U(r)\exp (ikz)$, where real function $U(r)$
obeys equation
\begin{equation}
kU-\frac{1}{2}\nabla ^{2}U+[g_{0}+g(\mathbf{r})]U^{5}=0,  \label{Stationary}
\end{equation}%
that can be derived from the Lagrangian density
\begin{equation}
2\mathcal{L}=kU^{2}+\frac{1}{2}\left( \nabla U\right) ^{2}+\frac{1}{3}\left[
g_{0}+g(\mathbf{r})\right] U^{6}.  \label{L}
\end{equation}%
In the 1D case, Eq. (\ref{Stationary}) taken at the inflexion point, $%
U^{\prime \prime }=0$, demonstrates that all solitons may exist only with $%
k<0$. A similar argument readily predicts $k<0$ for 2D solitons. With regard
to this, the Thomas-Fermi approximation (TFA), which neglects the
diffraction term, $\nabla ^{2}U$, yields a solution
\begin{equation}
U_{\mathrm{TFA}}^{2}(\mathbf{r})\approx \sqrt{-k/g(\mathbf{r})}  \label{TFA}
\end{equation}%
at $r\rightarrow \infty $, hence, as mentioned above, the modes with a
finite norm, $N_{\mathrm{2D}}=\int \int U^{2}(\mathbf{r})dxdy$, or $N_{%
\mathrm{1D}}=\int_{-\infty }^{+\infty }U^{2}(x)dx$, exist if $g(\mathbf{r})$
grows faster than $r^{2D}$ at $r\rightarrow \infty $.

In the present work, the 1D modulation function is taken as%
\begin{equation}
g(x)=\sinh ^{2}(\beta x),  \label{g}
\end{equation}%
the coefficient in front of which is scaled to be $1$, while $g_{0}$ is kept
as a free parameter in Eq. (\ref{GPE}). In fact, the unlimited exponential
growth of the nonlinearity strength at $|x|\rightarrow \infty $, which may
be difficult to realize in real physical settings, is not necessary for the
creation of the solitons supported by the spatially modulated SDF
nonlinearity. Indeed, the strong localization of the solitons predicted by
expressions (\ref{exact}) and (\ref{tail}) (see below) implies that the
modulation profile must actually represent a deep nonlinear-potential well
of a finite extension, rather than the unlimitedly growing structure, as a
particular form of $g(x)$ at $|x|\gg \beta ^{-1}$ does not affect the
solution.

In the case of $g_{0}<1$, Eq. (\ref{Stationary}) with $g(x)$ taken as per
Eq. (\ref{g}) admits an exact soliton at a particular value of $k$:
\begin{eqnarray}
U^{2}\left( x\right)  &=&\sqrt{\frac{3\beta ^{2}}{8\left( 1-g_{0}\right) }}%
\mathrm{sech}\left( \beta x\right) ,  \label{exact} \\
k &=&-\frac{\beta ^{2}}{8}\frac{2+g_{0}}{1-g_{0}}.  \label{k}
\end{eqnarray}%
It is worthy to mention that a particular exact solution can be also found
in the 1D model with a more general nonlinearity-modulation function:
\begin{equation}
iu_{z}=-\frac{1}{2}u_{xx}+\left[ \cosh \left( \beta x\right) \right] ^{\mu
}[g_{0}+\sinh ^{2}\left( \beta x\right) ]|u|^{4}u,  \label{mu}
\end{equation}%
where the real power takes values $\mu >-2$. In fact, most interesting are
the negative values, $-2<\mu <0$, as in that case the growth of the
nonlinearity at $|x|\rightarrow \infty $ is \emph{slower}. The exact soliton
solution found in this model is%
\begin{eqnarray}
u &=&e^{ikz}A\left[ \mathrm{sech}\left( \beta x\right) \right] ^{\left(
2+\mu \right) /4},  \label{u} \\
A^{4} &=&\frac{\beta ^{2}\left( 2+\mu \right) \left( 6+\mu \right) }{%
32\left( 1-g_{0}\right) },~  \label{A} \\
k &=&-\frac{\beta ^{2}\left( 2+\mu \right) }{32}\left[ \frac{6+\mu }{1-g_{0}}%
-\left( 2+\mu \right) \right] .  \notag
\end{eqnarray}%
In the limit of $\mu \rightarrow -2$, when the nonlinearity modulation in
Eq. (\ref{mu}) ceases to be growing at $|x|\rightarrow \infty $, expression (%
\ref{A}) for the amplitude degenerates into $A=0$. On the other hand, in the
limit of $2+\mu \approx C\left( 1-g_{0}\right) \rightarrow 0$ with fixed $C>0
$, Eq. (\ref{mu}) degenerates into the NLS equation with the constant
coefficient of the SDF quintic nonlinearity. In this case, the soliton goes
over into the continuous-wave (CW) solution with the constant amplitude, $A_{%
\mathrm{CW}}^{4}=\beta ^{2}C/8$. It is obvious that the latter
solution is modulationally stable \cite{ref5}, hence the continuity
suggests that the fundamental soliton solutions of Eq. (\ref{mu})
may be stable too, which is confirmed below (for $\mu =0$).

Coming back to general soliton solutions to the 1D version of Eq. (\ref{GPE}%
) with modulation function (\ref{g}), it is easy to find the asymptotic form
of the soliton solutions at $|x|\rightarrow \infty $, which accounts for its
strong localization:%
\begin{equation}
U(x)\approx \left( \frac{1}{2}\beta ^{2}-4k\right) ^{1/4}\exp \left( -\frac{1%
}{2}\beta |x|\right)   \label{tail}
\end{equation}%
(note that, unlike the usual solitons, the localization length determined by
this expression, $\sim \beta ^{-1}$, does not depend on $k$). Further, the
TFA applies to the entire 1D solution in the case of $-8k\gg \beta ^{2}$,
predicting the solution in the form of Eq. (\ref{TFA}) at all $x$, the
respective norm being%
\begin{equation}
N_{\mathrm{TFA}}\approx \left( 2I/\beta \right) \sqrt{-k},  \label{NTFA}
\end{equation}%
with constant $I\equiv \int_{0}^{\infty }\left( 2g_{0}-1+\cosh y\right)
^{-1/2}dy$. It is obvious that the fundamental (nodeless) solitons are
stable within the framework of the TFA.

It may be conjectured that the fundamental solitons realize the ground state
in the systems of the present type \cite{ref13}. A ``naive" counter-argument
against this assumption is a proposal to spread out the given norm, $N$,
into a layer of an indefinitely large length, $L$, with a vanishingly small
squared amplitude, $N/L$, so that the energy of such a state would fall to
zero, along with its amplitude, while the total energy of the fundamental
solitons is, obviously, positive. However, a straightforward estimate of the
energy of the stretching layer [the energy density is actually the same as
Lagrangian density (\ref{L}), except for the first term in it] yields an
estimate, $E(L)\approx \left( N^{3}/12\beta L^{3}\right) \exp \left( 2\beta
L\right) $, which \emph{diverges}, rather than vanishing, in the limit of $%
L\rightarrow \infty $.

A curious fact is that expression (\ref{tail}) gives an exact solution for
the \emph{entire family} of fundamental solitons in the model with $%
g(x)=(1/4)\exp \left( 2\beta |x|\right) $, which emulates the asymptotic
form of function (\ref{g}), if, in addition to the effective nonlinear
potential, an attractive linear delta-functional potential is placed at $x=0$%
. The respective stationary equation is
\begin{equation}
kU-\frac{1}{2}U^{\prime \prime }+\frac{1}{4}e^{2\beta |x|}U^{5}-\frac{\beta
}{2}\delta (x)U=0.  \label{concocted}
\end{equation}%
The linear potential $-\left( \beta /2\right) \delta (x)$ in Eq. (\ref%
{concocted}) is necessary to balance the peakon singularity in expression (%
\ref{tail}), if it is considered as the exact solution. Obviously, the norm
of this solution is $N(k)=\beta ^{-1}\sqrt{2\left( \beta ^{2}-8k\right)
\text{,}}$ with the propagation constant taking values $k\leq \beta ^{2}/8$
(note that this $k$ may be positive too).

Furthermore, the spatially modulated SDF nonlinearity can trap a soliton
against the action of the linear \emph{repulsive} potential. To demonstrate
this possibility, one can take the following modification of Eqs. (\ref%
{Stationary}) and (\ref{g}),%
\begin{equation}
kU-\frac{1}{2}\nabla ^{2}U+[g_{0}+\sinh ^{2}\left( \beta x\right) ]U^{5}+W%
\mathrm{sech}^{2}\left( \beta x\right) U=0,  \label{W}
\end{equation}%
where $W>0$ is the strength of the repulsive linear potential. An exact
solution to Eq. (\ref{W}) for the trapped mode can be found in the following
form, which generalizes the above solution given by Eqs. (\ref{exact}) and (%
\ref{k}):%
\begin{eqnarray}
U^{2}\left( x\right) &=&\sqrt{\frac{1}{1-g_{0}}\left( \frac{3\beta ^{2}}{8}%
+W\right) }\mathrm{sech}\left( \beta x\right) , \\
k &=&-\frac{1}{1-g_{0}}\left[ \frac{\beta ^{2}}{8}\left( 2+g_{0}\right) +W%
\right] .
\end{eqnarray}%
Numerical results [not shown here in detail, but essentially the same as
presented below, i.e., based on the analysis of small perturbations---see
Eq. (\ref{perturb})---and direct simulations] demonstrate that this solution
is stable.

\subsection{The variational approximation}

\subsubsection{The Gaussian ansatz}

To search for soliton solutions of Eq. (\ref{Stationary}) by means of the
VA, we start with the simplest Gaussian ansatz,
\begin{equation}
U^{2}\left( x\right) =\frac{N}{\sqrt{\pi }W}\exp \left( -\frac{x^{2}}{2W^{2}}%
\right) ,  \label{Gauss}
\end{equation}%
where $W$ and $N$ are the width and norm of the soliton. The substitution of
this ansatz into Lagrangian density (\ref{L}) and subsequent integration
yields the total Lagrangian, $L=\int_{-\infty }^{+\infty }\mathcal{L}dx$,
from which it is straightforward to derive the variational equations, $%
\partial L_{\mathrm{eff}}/\partial N=\partial L_{\mathrm{eff}}/\partial W=0$%
:
\begin{equation}
\frac{1}{4W^{2}}+\frac{N^{2}}{2\sqrt{3}\pi W^{2}}\left[ 2g_{0}+1-e^{-\frac{%
(\beta W)^{2}}{3}}\right] =-k,  \label{GW}
\end{equation}%
\begin{equation}
\frac{1}{2W^{2}}+\frac{N^{2}}{3\sqrt{3}\pi }\left[ \frac{(2g_{0}+1)}{W^{2}}%
+\left( \frac{\beta ^{2}}{3}-\frac{1}{W^{2}}\right) e^{\frac{(\beta W)^{2}}{3%
}}\right] =0.  \label{GN}
\end{equation}

\subsubsection{The raised-sech ansatz}

The availability of the particular exact solution (\ref{exact}), expressed
in terms of $\mathrm{sech}$, suggests to use this function as an alternative
variational ansatz, as it is natural to have one which is able to reproduce
a particular exact solution, suggesting a better accuracy in the general
case too (which turns out to be correct, see below). Usually, the ansatz
based on $\mathrm{sech}$ is introduced with arbitrary amplitude and width
\cite{VA}. However, in the present model the functional form of the ansatz
must be fixed as $\mathrm{sech}\left( \beta x\right) $, with the same $\beta
$ as in Eq.(\ref{g}), as otherwise the integration of the term in the
Lagrangian density (\ref{L})\ containing $g(x)$ is impossible in an
analytical form. The arbitrary width can be accommodated differently,
adopting the following \textit{raised-sech} ansatz (sech raised to arbitrary
power $\nu >0$):%
\begin{equation}
U(x)=A[\mathrm{sech}\left( \beta x\right) ]^{\nu },  \label{U}
\end{equation}%
where $A$ and $\nu \ $are to be treated as variational parameters. In
particular, in the case of $\nu \ll 1$, the standard FWHM width of ansatz (%
\ref{U}) is $\left( \ln 2\right) /\left( \beta \nu \right) $, which
demonstrates how $\nu $ controls the width.

To the best of our knowledge, the VA based on ansatz (\ref{U}), with power $%
\nu $ dealt with as the variational parameter, is a technical novelty, which
may be helpful too in studies of other models with the width of
shape-defining function fixed by the form of the given equation(s). As shown
below, this ansatz leads to rather complex variational equations which,
nevertheless, can be solved, leading to a good agreement with directly found
numerical results. It is relevant to note that some still more sophisticated
trial functions were recently proposed for the description of solitons, such
as the so-called $q$-Gaussian ansatz \cite{q}, with extra an parameter ($q$)
controlling a transition between the limit cases of the Gaussian and TFA
\textit{ans\"{a}tze}.

The effective Lagrangian, produced by the integration of density (\ref{L})
with ansatz (\ref{U}) substituted into it, is%
\begin{equation}
\frac{2}{\sqrt{\pi }\beta }L=\left[ k+\frac{\nu ^{2}\beta ^{2}}{2\left( 2\nu
+1\right) }\right] \frac{A^{2}\Gamma \left( \nu \right) }{\Gamma \left( \nu +%
\frac{1}{2}\right) }+\frac{A^{6}}{3}\frac{g_{0}\Gamma \left( 3\nu \right) }{%
\Gamma \left( 3\nu +\frac{1}{2}\right) }+\frac{A^{6}}{3\left( 6\nu -1\right)
}\frac{\Gamma \left( 3\nu -1\right) }{\Gamma \left( 3\nu -\frac{1}{2}\right)
},  \label{LL}
\end{equation}%
where $\Gamma $ is the Gamma-function (below, $\Gamma ^{\prime }$ stands for
its derivative). The corresponding variational equations, $\partial
L/\partial \left( A^{2}\right) =0,\ \partial L/\partial \nu =0,$ take the
following form:
\begin{equation}
\left[ k+\frac{\nu ^{2}\beta ^{2}}{2\left( 2\nu +1\right) }\right] \frac{%
\Gamma \left( \nu \right) }{\Gamma \left( \nu +\frac{1}{2}\right) }+\frac{%
g_{0}A^{4}\Gamma \left( 3\nu \right) }{\Gamma (3\nu +\frac{1}{2})}+\frac{%
A^{4}}{6\nu -1}\frac{\Gamma (3\nu -1)}{\Gamma (3\nu -\frac{1}{2})}=0,
\label{NewVAa}
\end{equation}%
\begin{gather}
\left[ k+\frac{\nu ^{2}\beta ^{2}}{2\left( 2\nu +1\right) }\right] \left[
\frac{\Gamma ^{\prime }(\nu )}{\Gamma (\nu +\frac{1}{2})}-\frac{\Gamma (\nu
)\Gamma ^{\prime }(\nu +\frac{1}{2})}{\Gamma ^{2}(\nu +\frac{1}{2})}\right] +%
\frac{\beta ^{2}\nu (\nu +1)}{\left( 2\nu +1\right) ^{2}}\frac{\Gamma (\nu )%
}{\Gamma (\nu +\frac{1}{2})}+  \notag \\
\frac{g_{0}A^{4}}{3}\left[ \frac{\Gamma ^{\prime }\left( 3\nu \right) }{%
\Gamma (3\nu +\frac{1}{2})}-\frac{\Gamma \left( 3\nu \right) \Gamma ^{\prime
}(3\nu +\frac{1}{2})}{\Gamma ^{2}(3\nu +\frac{1}{2})}\right] +\frac{A^{4}}{%
3\left( 6\nu -1\right) }\left[ \frac{\Gamma ^{\prime }(3\nu -1)}{\Gamma
(3\nu -\frac{1}{2})}-\frac{\Gamma (3\nu -1)\Gamma ^{\prime }(3\nu -\frac{1}{2%
})}{\Gamma ^{2}(3\nu -\frac{1}{2})}\right] =0.  \label{NewVAb}
\end{gather}%
In spite of the relative complexity of these equations, it is possible to
check that, setting $\nu =1/2$ and $g_{0}=0$, they reproduce the
corresponding exact solution (\ref{exact}) with $k$ given by Eq. (\ref{k}).

\begin{figure}[tbp]
\begin{center}
\includegraphics[height=4.cm]{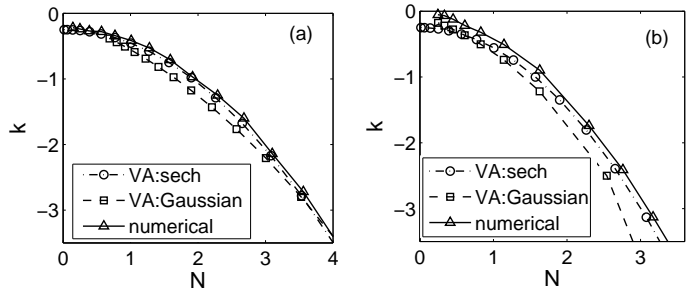}
\end{center}
\caption{The comparison of predictions of the VA for fundamental solitons,
based on the Gaussian and raised-$\mathrm{sech}$ \textit{ans\"{a}tze}, with
numerically found dependences $k(N)$, for $\protect\beta =2$ and $g_{0}=0$
(a) and $g_{0}=0.5$ (b) . }
\label{Fig1}
\end{figure}

The comparison between both versions of the VA, produced by numerical
solutions of Eqs. (\ref{GW}), (\ref{GN}) and (\ref{NewVAa}), (\ref{NewVAb}),
respectively, and results obtained from a numerical solution of Eq. (\ref%
{Stationary}) are presented in Fig. \ref{Fig1}. It is observed that ansatz (%
\ref{U}) provides for a substantial improvement of the accuracy of the VA
[expect for the limit case of small $N$ in Fig. \ref{Fig1}(b)]. It is
interesting to note that, according to the variational and numerical results
alike, the entire family of the fundamental solitons satisfies the
above-mentioned \textit{anti-VK} criterion, $dk/dN<0$, which was recently
proposed, at a semi-empiric level, as a stability criterion for fundamental
bright solitons in media with repulsive nonlinearities \cite{antiVK,ref12}.
Indeed, it is demonstrated below that the fundamental solitons are fully
stable in the present model.

It is also worthy to stress that, as shown by the $k(N)$ curves (the
numerical and VA-predicted ones alike) in Fig. \ref{Fig1}, the family of the
fundamental solitons features no existence threshold, i.e., the solutions
persist up to the limit of $N\rightarrow 0$. An analytical consideration of
Eqs. (\ref{Stationary}) and (\ref{g}) suggests that, in this limit, the
soliton acquires a shape of a quasi-flat shelf of width $L\simeq \left(
2/\beta \right) \ln \left( \beta /N\right) $, with the squared amplitude $%
A^{2}\simeq N/L$.

\section{Numerical results}

\subsection{The linear stability analysis}

It is crucially important to test stability of the solitons. To this end,
perturbed solutions were taken as
\begin{equation}
u(x,z)=\exp (ikz)[U(x)+V(x)\exp (\lambda z)+W^{\ast }(x)\exp (\lambda ^{\ast
}z)],  \label{perturb}
\end{equation}%
where $V$ and $W$ are components of a perturbation mode with growth rate $%
\lambda $, while $U(x)$ is the stationary solution to Eq. (\ref{GPE}) with
propagation constant $k$ (the asterisk stands for the complex conjugate).
The substitution of expression (\ref{perturb}) into Eq. (\ref{GPE}) and
subsequent linearization gives rise to the eigenvalue problem for $\lambda $%
:
\begin{equation}
i\lambda V=-\left( 1/2\right) V_{xx}+kV+[g_{0}+g(x)]U^{4}(3V+2W),
\label{Eigena}
\end{equation}%
\begin{equation}
i\lambda W=+\left( 1/2\right) W_{xx}-kW-[g_{0}+g(x)]U^{4}(3W+2V).
\label{Eigenb}
\end{equation}%
Obviously, the stationary solution, $U(x)$, is stable only if $\mathrm{Re}%
\left\{ \lambda \right\} =0$ for all the eigenvalues.

The eigenvalue problem can be solved numerically by means of a
finite-difference scheme. Specifically, we first took the Gaussian ansatz as
the initial guess to construct solutions $U(x)$ of Eq. (\ref{Stationary}),
and then tested the stability of this so found solutions by solving the
eigenvalue problem based on Eqs. (\ref{Eigena}) and (\ref{Eigenb}). Finally,
the predicted (in)stability was tested in direct simulations of Eq. (\ref%
{GPE}). The numerical computations were performed in the domain of size $%
-10\leq x\leq 10$ on a grid of 128 points, with the usual absorbing
boundary conditions. This method is reliable, as it has been applied
extensively to soliton-related problems. For further details, see
book \cite{Eigenvalue}, which contains the relevant code. Results
produced by the linear-stability analysis are presented below.

\subsection{Fundamental and higher-order solitons}

Examples of fundamental (nodeless) and higher-order solitons found in the
numerical form are displayed in Figs. 2(a) and 2(b,c), respectively. Note
that the raised-$\mathrm{sech}$ ansatz predicts the profile of the
fundamental soliton which is virtually identical to its numerically found
counterpart. Both the computation of the eigenvalues for small perturbations
and direct simulations demonstrate that the entire family of the fundamental
solitons is stable.

Higher-order solitons are characterized by the number of nodes (zeros of the
field). In particular, the examples shown in Fig. 2(b,c) feature two zero
crossings, and examples of dipole solitons, with the single node, are
presented in Fig. \ref{Fig3}.

\begin{figure}[tbp]
\begin{center}
\includegraphics[height=7.cm]{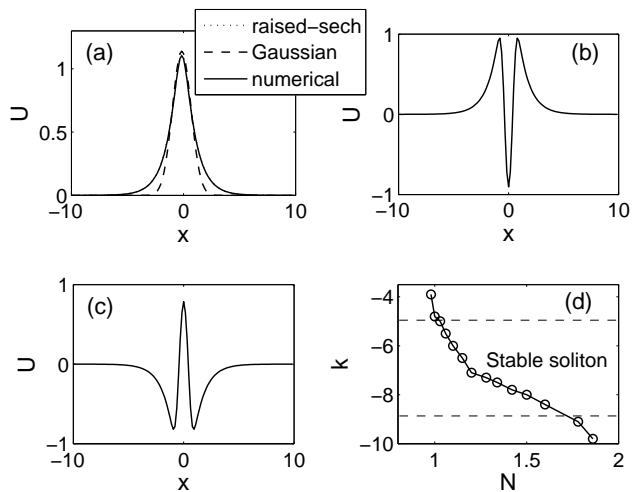}
\end{center}
\caption{(a) An example of a stable 1D fundamental soliton, and the
comparison with the Gaussian and raised-$\mathrm{sech}$ \textit{ans\"{a}tze}%
. The latter one is indistinguishable from the numerically found profile.
(b,c) Examples of stable higher-order solitons with two nodes. These
examples are displayed for $\protect\beta =2$ and $g_{0}=0$, other
parameters being (a) $k=-1$, $N=1.91$, (b) $k=-8.65$, $N=1.6$ , and (c) $%
k=-7.1$, $N=1.2$. (d) The propagation constant vs. the norm for the
higher-order solitons with two nodes at $\protect\beta =2$ and $g_{0}=0$.
The solitons are stable between the dashed lines. }
\label{Fig2}
\end{figure}

\begin{figure}[tbp]
\begin{center}
\includegraphics[height=4.cm]{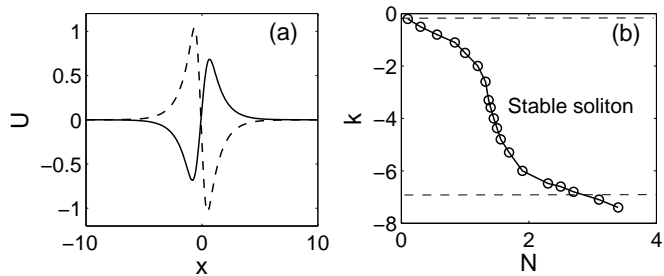}
\end{center}
\caption{(a) Examples of stable dipole solitons for $\protect\beta =2$ and $%
g_{0}=0$. Other parameters are $k=-2.6$, $N=1.32$ and $k=-4.8$, $N=1.56$
(the dashed and solid profiles, respectively). (b) Curve $k(N)$ for the
dipole solitons at $\protect\beta =2$ and $g_{0}=0$. The solitons are stable
between the dashed lines.}
\label{Fig3}
\end{figure}

The solitons with two nodes, as well as the dipole solitons, are stable in a
part of their existence regions, as shown in Fig. 2(d) and 3(b),
respectively. Note that the entire $k(N)$ curves in these figures satisfy
the anti-VK criterion, $dk/k{N}<0$, as well as they did for the fundamental
solitons, cf. Fig. \ref{Fig1}. The fact that only a part of the families of
the higher-order solitons is stable complies with the general fact that the
usual VK criterion, in models with self-focusing nonlinearities, is
necessary but not sufficient for the full stability of bright solitons \cite%
{VK,ref5}. Further, the conclusion that the fundamental solitons supported
by the SDF nonlinearity with the local coefficient growing at $r\rightarrow
\infty $ are completely stable, while a part of higher-order families are
unstable, agrees with the results recently reported in Refs. \cite{ref13}
and \cite{ref13-2} for the cubic nonlinearity with similar modulation
profiles. In fact, it was found that, in the case of the steep
(anti-Gaussian) modulation, the solitons with one and two nodes are
completely stable, instability regions appearing for modes with three zeros
\cite{ref13}, while, under the action of a milder algebraic modulation, with
$g(x)\sim |x|^{\alpha }$ ($\alpha >1$), the solitons with one and two zeros
may be unstable too, only the fundamental family remaining completely
stable. The exponential modulation function adopted in the present model,
see Eq. (\ref{g}), is intermediate between its steep anti-Gaussian and mild
algebraic counterparts.

\begin{figure}[tbp]
\begin{center}
\includegraphics[height=7.cm]{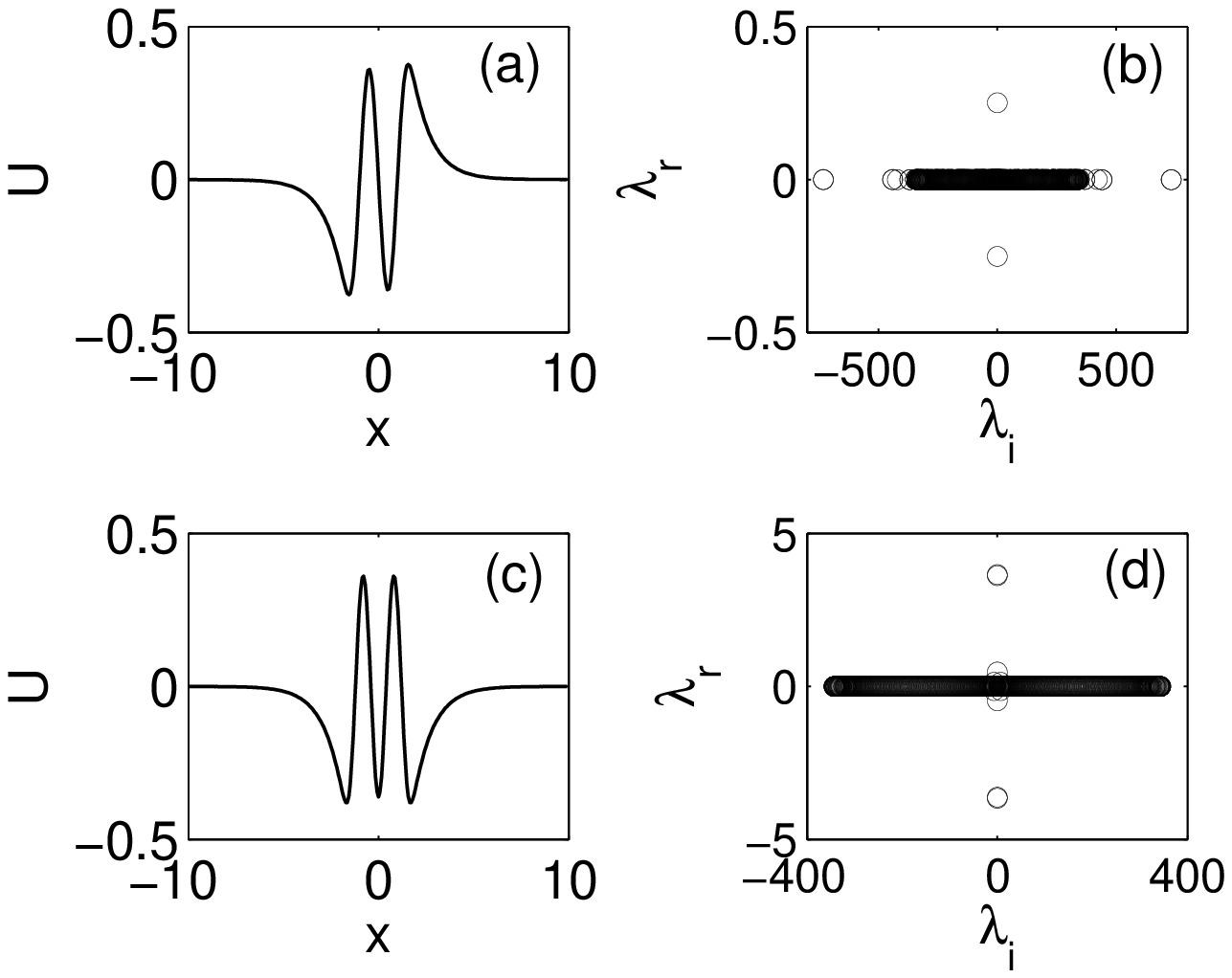}
\end{center}
\caption{The shapes (left) and their perturbation spectra (right) of
unstable 1D modes with three (a,b) and four (c,d) zeros, at $\protect\beta %
=2 $, $g_{0}=0$. (a,b) $k=-5$, $N=1.03$; (c,d) $k=-7.6$, $N=1.07$.}
\label{Fig4}
\end{figure}

\begin{figure}[tbp]
\begin{center}
\includegraphics[height=6.cm]{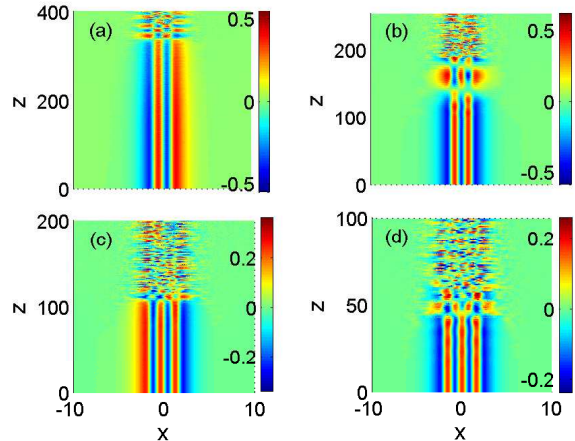}
\end{center}
\caption{(Color online) The evolution, shown in terms of $\mathrm{Re}\left\{
u(x,z)\right\} $, of unstable 1D higher-order solitons with the number of
nodes $n=3,4,5,$ and $6$ at $\protect\beta =2$, $g_{0}=0$. (a) $k=-5.3$, $%
N=1.05$; (b) $k=-8.9$, $N=1.35$; (c) $k=-7.6$, $N=1.07$; (d) $k=-7.7$, and $%
N=1.2$.}
\label{Fig5}
\end{figure}

We have found that all the higher-orders modes with $\geq 3$ zeros are
unstable in the present model, as illustrated by Figs. \ref{Fig4} and \ref%
{Fig5}. The instability transforms the unstable solitons into chaotically
oscillating localized modes, see examples in Fig. \ref{Fig5}.

\section{Solitons in the 2D model}

An issue of obvious interest is to extend the model and its analysis to the
2D geometry, cf. Refs. \cite{ref13} and \cite{ref13-2}. Although the TG
model is irrelevant in 2D, the above-mentioned optical realization, in terms
of the colloidal suspensions, applies to the 2D case too. To produce an
example of exact 2D solutions, we set $g_{0}=0$ in Eq. (\ref{GPE}) and take
\begin{equation}
g(x,y)=g_{0}+[\beta ^{-2}\sinh ^{2}(\beta x)\cosh ^{2}(\alpha y)+\alpha
^{-2}\sinh ^{2}(\alpha y)\cosh ^{2}(\beta x)],  \label{g2D}
\end{equation}%
cf. Eq. (\ref{g}) in 1D. In this case, the following exact fundamental
soliton can be found, for $g_{0}=0$: $u\left( x,y,t\right) =U(x,y)\exp (ikz)$%
, with $k=-(\alpha ^{2}+\beta ^{2})/4$ and
\begin{equation}
U^{2}\left( x,y\right) =\sqrt{3/8}\alpha \beta ~\mathrm{sech}\left( \beta
x\right) \mathrm{sech}\left( \alpha y\right)  \label{exact2D}
\end{equation}

Examples of isotropic $\left( \alpha =\beta =1\right) $ and anisotropic $%
\left( \alpha =1,\beta =1/2\right) $ 2D solitons, found in the numerical
form for modulation function (\ref{g2D}) with $g_{0}=0.1$, when exact
solutions are not available, are shown in Fig. 6. The stability of these
solitons was verified by direct simulations of the perturbed evolution.
Systematic results for 2D solitons will be reported elsewhere.

\begin{figure}[tbp]
\begin{center}
\includegraphics[height=6.cm]{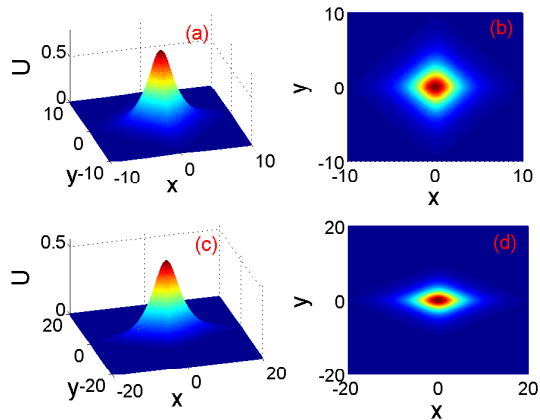}
\end{center}
\caption{(Color online) Shapes of stable isotropic and anisotropic 2D
fundamental solitons found from Eqs. (\protect\ref{Stationary}) and (\protect
\ref{g2D}) with $g_{0}=0.1$. (a, b) $\protect\alpha =\protect\beta =1$ and $%
k=-0.5$, $N=4$; (c, d) $\protect\alpha =2\protect\beta =1$ and $k=-0.3125$, $%
N=3.5$.}
\label{Fig6}
\end{figure}

\section{Conclusion}

We have introduced the 1D and 2D models with the SDF (self-defocusing)
quintic nonlinearity growing at $r\rightarrow \infty $. The model may be
realized in optical media, and its 1D version can be also implemented in the
Tonks-Girardeau gas. Recently, stable bright solitons were found in similar
models with the cubic nonlinearity. The objective of the present work is to
test the genericity of the mechanism creating bright solitons in cubic SDF
media with the spatially modulated nonlinearity coefficient, by testing it
with the other (quintic) nonlinearity. We have found particular exact
solutions for fundamental solitons in the 1D model with the modulation
function defined as per Eqs. (\ref{Stationary}) and (\ref{g}). General
solutions have been found in the numerical form, and also analytically in
the framework of the TFA and VA (Thomas-Fermi and variational
approximations). In particular, a new ansatz for the VA, based on the raised
$\mathrm{sech}$, was developed. It yields an essentially better accuracy
than the usual Gaussian ansatz. All the fundamental solitons are stable,
while higher-order ones have a finite stability region for modes with one
and two nodes, all the solitons with $\geq 3$ nodes being unstable. It has
been found that the recently proposed ``anti-VK" stability criterion for
bright solitons in SDF media is valid, as a necessary stability condition,
in the present model too. The 2D model was also considered, in a brief form.
Particular exact solutions for 2D solitons were produced, and examples of
numerically found stable solitons in 2D were reported, both isotropic and
anisotropic ones. The analysis of the 2D model calls for an extension; in
particular, a challenging problem is to construct solutions for vortex
solitons.

\section{Acknowledgment}

   J.Z. acknowledges support from the Natural Science Foundation of China
(project No. 11204151). The work of B.A.M. was supported, in a part, by the German-Israel Foundation (Grant No. I-1024-2.7/2009).

\end{document}